\documentclass{INTERSPEECH2023}
\usepackage{amsmath,graphicx}

\def\h{{\mathbf h}}

\def\z{{\mathbf z}}

\def\etal{{\em et al.~}}

\usepackage{pifont}
\newcommand{\cmark}{\ding{51}}%
\usepackage{tabularx}
\usepackage{booktabs}
\usepackage{multirow}
\usepackage{ragged2e}
\usepackage{bm}

\usepackage{boldline}

\newcolumntype{C}{>{\Centering\arraybackslash}X}

\newcolumntype{u}{>{\raggedright\hsize=.7\hsize}X}
\newcolumntype{t}{>{\Centering\hsize=.6\hsize}X}
\newcolumntype{s}{>{\Centering\hsize=.5\hsize}X}
\newcolumntype{o}{>{\Centering\hsize=.4\hsize}X}
\newcolumntype{k}{>{\Centering\hsize=.3\hsize}X}
\newcolumntype{y}{>{\Centering\hsize=.2\hsize}X}
\newcolumntype{z}{>{\Centering\hsize=.1\hsize}X}
\newcolumntype{v}{>{\raggedright\hsize=.6\hsize}X}
\newcolumntype{e}{>{\raggedright\hsize=.5\hsize}X}
\newcolumntype{j}{>{\raggedright\hsize=.35\hsize}X}
\newcolumntype{f}{>{\raggedright\hsize=.3\hsize}X}
\newcolumntype{h}{>{\raggedright\hsize=.2\hsize}X}
\newcolumntype{q}{>{\raggedright\hsize=.8\hsize}X}

\usepackage[backend=bibtex,citestyle=numeric-comp,bibstyle=ieee,sorting=none,defernumbers=true,giveninits=true,doi=false,isbn=false,url=false,eprint=false,minbibnames=1,maxbibnames=5]{biblatex}
\interspeechcameraready

\title{Streaming Audio-Visual Speech Recognition with Alignment Regularization}
\name{Pingchuan Ma$^{1,2}$, Niko Moritz$^{1}$, Stavros Petridis$^{1,2}$, Christian Fuegen$^{1}$, Maja Pantic$^{1,2}$
\thanks{Only non-Meta co-authors downloaded, accessed, and used the datasets. Only non-Meta authors conducted any of the dataset pre-processing (no dataset pre-processing took place on Meta’s servers or facilities).}
}
\address{$^1$Meta AI\\$^2$Imperial College London}
\email{pingchuanma@meta.com, nmoritz@meta.com, stavrosp@meta.com}

\def\ninept{\def\baselinestretch{.95}\let\normalsize\small\normalsize}
\def\tenpt{\def\baselinestretch{1.1}\let\normalsize\large\normalsize}

\begin{document}
\ninept
\maketitle
 
\begin{abstract}
In this work, we propose a streaming AV-ASR system based on a hybrid connectionist temporal classification (CTC)/attention neural network architecture. The audio and the visual encoder neural networks are both based on the conformer architecture, which is made streamable using chunk-wise self-attention (CSA) and causal convolution. Streaming recognition with a decoder neural network is realized by using the triggered attention technique, which performs time-synchronous decoding with joint CTC/attention scoring. Additionally, we propose a novel alignment regularization technique that promotes synchronization of the audio and visual encoder, which in turn results in better word error rates (WERs) at all SNR levels for streaming and offline AV-ASR models. The proposed AV-ASR model achieves WERs of 2.0\,\% and 2.6\,\% on the Lip Reading Sentences 3 (LRS3) dataset in an offline and online setup, respectively, which both present state-of-the-art results when no external training data are used.
\end{abstract}

\section{Introduction}
\label{sec:intro}
Audio-Visual Automatic Speech Recognition (AV-ASR) can be defined as the task of recognizing speech by using both audio and visual information. It has recently attracted a lot of attention due to the fact that AV-ASR models allow for increased robustness to acoustic noise. However, the vast majority of works have focused on developing AV-ASR models for non-streaming recognition. For example, \cite{afouras2018deep, petridis2018audio, DBLP:journals/corr/abs-2102-06657, serdyuk2021audiovisual} built AV-ASR models by using either a transformer or conformer encoder on top of the acoustic and visual front-ends for speech recognition, followed by a fusion module that jointly learns the audio-visual encoder state sequence. Finally, a decoder network maps the encoder state sequences to output label sequence using the attention mechanism~\cite{watanabe2017hybrid, DBLP:conf/nips/VaswaniSPUJGKP17}. However, such attention-based encoder-decoder models are limited to offline scenarios since an entire speech utterance is typically required as an input. To the best of our knowledge, Taris~\cite{DBLP:journals/csl/SterpuH22} is the only AV-ASR model to perform streaming speech recognition but no results are reported on the largest publicly available audio-visual benchmarks.

Various works~\cite{DBLP:conf/icassp/YuZK21, DBLP:conf/icmi/SterpuSH18, sterpu2020teach} have shown that AV-ASR models do not always outperform audio-only models in clean scenarios~\cite{serdyuk2022transformer,DBLP:journals/corr/abs-2102-06657}. One reason for this is that AV-ASR models tend to ignore the visual cues in clean scenarios, where audio typically outperforms the visual modality. However, because the visual modality often demonstrates advantages in noisy situations, some works~\cite{DBLP:journals/corr/abs-2102-06657, DBLP:journals/corr/abs-2201-01763} propose to inject noise for the audio-visual training to encourage the model to better take the visual modality into account. Sterpu~\etal~\cite{DBLP:conf/icmi/SterpuSH18, sterpu2020teach} proposed a cross-modal attention mechanism to automatically learn better alignments between audio and visual streams, which also encourages the system to pay attention to both modalities. However, in a streaming system, where access to future frames is limited, techniques such as cross-modal attention, which performs temporal attention between two sequences, may be sub-optimal, because it would introduce an additional delay.

In this work, we propose a novel regularization technique that promotes synchronization of the audio and visual encoder outputs. In an AV-ASR system, audio and visual encoder neural networks are often pre-trained independently and outputs are fused by an audio-visual fusion operation that simply stacks frames of the encoder outputs from both modalities before being further processed by a neural network~\cite{makino2019recurrent, DBLP:journals/corr/abs-2102-06657}. However, there is no guarantee that responses of both encoders are aligned and the operation of stacking features fails to compensate for such misalignments. As a result, a frame-level ASR loss, such as CTC or recurrent neural network transducers (RNN-T)~\cite{GravesFGS06,Graves12}, would only be able to make decisions based on one of the modalities, because both encoders may respond to inputs at different frames. To mitigate this problem, we propose to use forced alignment information from the AV-ASR model and two auxiliary cross-entropy-based (CE) losses to better synchronize audio and visual encoder outputs. We can show that such regularization indeed leads to better frame-level alignments of separate audio and visual encoder streams, which in turn results in a more robust performance for our AV-ASR models across all noise conditions.

The proposed alignment regularization technique is applied for building offline and online AV-ASR systems, where the latter must recognize a word shortly after it is spoken. The model architecture 
is based on~\cite{DBLP:journals/corr/abs-2102-06657}, where the audio-only and visual-only conformer encoders are made streamable by using chunk-wise restricted self-attention (CSA) and causal convolution \cite{ChenWW21_streamingASR,Moritz21_DCN}. We also use the idea of triggered attention, which leverages the alignment information of the CTC output to trigger a attention-based decoder with a restricted look-ahead onto the encoder sequence \cite{DBLP:conf/icassp/MoritzHR20}.

\section{Model Setup}
\noindent\textbf{Architecture}\quad
\label{ssec:architecture}
\begin{figure}[!t]
    \centering
    \includegraphics[width=.87\columnwidth]{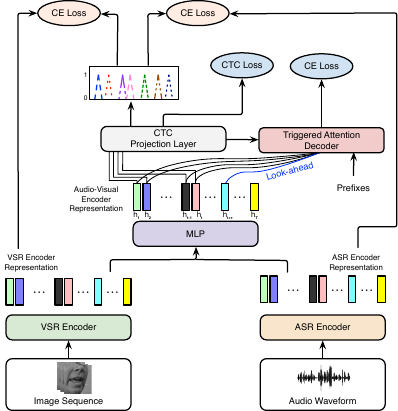}
    \caption{Triggered attention-based streaming AV-ASR model with alignment regularization, which uses the alignment information from the joint audio-visual CTC output to promote temporally aligned responses of the VSR and ASR encoders through auxiliary CE losses. $t$ is the time index for a current trigger position and $\tau$ denotes the look-ahead hyper-parameter of the TA decoder.
    }
    \label{fig:architecture}
    \vspace{-8mm}
\end{figure}
Our models are built based on the AV-ASR architecture proposed in~\cite{DBLP:journals/corr/abs-2102-06657}. The VSR encoder consists of a modified 2D ResNet-18 backbone as the visual front-end~\cite{DBLP:conf/icassp/PetridisSMCTP18}, followed by a 12-layer conformer encoder, whereas the ASR encoder consists of 1D ResNet-18 as the audio front-end and a 12-layer conformer encoder. The frame rate at the output of the 1D ResNet-18 is sub-sampled to 25 frames per second (fps), which matches the temporal resolution of the visual stream. The Conformer encoder is made streamable by using chunk-wise self-attention with non-overlapping chunks in order to not increase the computational complexity, which has also demonstrated superior results compared to restricted self-attention \cite{ChenWW21_streamingASR}. In the streaming setup, causal convolution is applied for the convolutional modules of the conformer architecture \cite{Moritz21_DCN}. For fusing the audio and visual encoder outputs, a 2-layer multi-layer perceptron (MLP) is used. The joint audio-visual representations, which are computed by the MLP module, are further fed to both CTC and a triggered attention decoder for the joint training. The architecture of the proposed AV-ASR model is illustrated in Fig.~\ref{fig:architecture}.

\noindent\textbf{Triggered attention}\quad
A set of solutions~\cite{DBLP:conf/iclr/ChiuR18, DBLP:conf/icassp/MoritzHR19} have been proposed to enable the application of streaming attention-based encoder-decoder models. Triggered attention (TA)~\cite{DBLP:conf/icassp/MoritzHR19, moritz2019streaming, DBLP:conf/icassp/MoritzHR20}, as one of the most popular and successful methods, uses the alignment properties of CTC to determine time positions at which an ASR output is generated and then triggers a attention-based decoder to perform attention over past encoder frames up to the trigger frame position plus some look-ahead encoder frames.
During training, such trigger positions can be determined using CTC-based forced alignment to train the decoder with limited encoder inputs~\cite{DBLP:conf/icassp/MoritzHR19}.
However, in this work we did not apply TA for training, since training with limited context did not help to improve ASR results when performing TA decoding. Thus, for simplicity, the decoder is trained in full-sequence mode and TA is only activated at inference time for streaming recognition.
At inference time, the decoding procedure follows the frame-synchronous TA decoding algorithm proposed by~\cite{moritz2019streaming, DBLP:conf/icassp/MoritzHR20}.
TA decoding leverages frame-synchronous CTC beam search to determine frame/time positions at which a new ASR token is detected to trigger the attention decoder. The TA attention decoder receives as an input all the past encoder frames up to the trigger frame plus some additional look-ahead frames. This generates additional ASR scores that are used for the joint CTC/attention scoring of ASR hypothesis.

\section{Alignment regularization}
Frame-level sequence-to-sequence ASR losses, such as CTC and RNN-T, learn an alignment correspondence between an input sequence, e.g., a sequence of acoustic or visual features, and a typically shorter output sequence, the ASR labels. This is achieved by allowing the system to output blanks at frames at which no ASR label can be detected. However, there are various ways where to output blanks or labels, which each denote a different alignment. CTC and RNN-T solve this problem by optimizing the sum of all possible alignments, which results in emphasizing one alignment, because alignments are mutually exclusive. However, such learned alignments are not deterministic and also depend on factors such as the neural network architecture or the kind of input features. This fact can be problematic when combining different modalities from separate encoder streams, because responses may not be aligned, even though audio-visual streams are synchronous at the input, which hinders the model to make joint decisions at frame-level.
Sanabria~\etal~\cite{DBLP:journals/corr/SanabriaMT16} provide an analysis of the response offset for each modality to demonstrate that their proposed audio-visual ASR systems reduce the response offset between audio and visual encoder state sequences.

In this work, we propose to use the alignments derived from trained CTC outputs and auxiliary CE losses to synchronize the audio and visual streams, where the best alignment path is determined using the Viterbi algorithm~\cite{DBLP:conf/icassp/MoritzHR19}.
It is worth pointing out that our approach allows alignment information to be extracted on-the-fly without generating alignments in advance from a hybrid ASR system.
Thus, the proposed alignment regularization technique aims to temporally synchronize the emission of label information by the audio and visual encoder neural networks, which in turn helps to increase robustness of the AV-ASR system.
We train our model using the following loss

\vspace{-3mm}
\begin{align}
\mathcal{L}=\alpha\mathcal{L}_{\text{\tiny CTC}}+(1-\alpha)\mathcal{L}_{\text{\tiny CE}}
+\mathcal{L}_{\text{\tiny ALIGN}}
\label{eq:trainingCTCweight}
\end{align}

where $\alpha$ denotes the relative weight to control the contribution between the CTC loss $\mathcal{L}_{\text{\tiny CTC}}$ and the attention decoder loss $\mathcal{L}_{\text{\tiny CE}}$. $\mathcal{L}_\text{\tiny ALIGN}$ represents an auxiliary loss for the alignment regularization, which is computed as follows:

\begin{align}
&\mathcal{L}_{\text{\tiny ALIGN}}=
\beta_{\text{\tiny AV2V}}\text{log} p_{\text{\tiny CE}}^{\text{\tiny AV2V}}(\z^{AV}|\h^{V})
+
\beta_{\text{\tiny AV2A}}\text{log} p_{\text{\tiny CE}}^{\text{\tiny AV2A}}(\z^{AV}|\h^{A})
\label{eq:alignment loss}
\end{align}

In Eq.~\ref{eq:alignment loss}, $\h^{A}=(h^{A}_1, ..., h^{A}_T)$ and $\h^{V}=(h^{V}_1, ..., h^{V}_T)$ denote the ASR and VSR encoder output sequences, respectively, with $T$ being the number of encoder frames. The frame-level audio and visual encoder representations are denoted as $h^A$ (audio) and $h^V$ (visual).
$\z^{AV}=(z^{AV}_1, ..., z^{AV}_T)$ denotes the best alignment path that is derived by CTC-based forced alignment using the output of the CTC projection layer for the joint audio-visual encoder and the ground truth label sequence, where each $z^{AV}_t$ represent a label such as a blank or non-blank symbol with $t$ indicating the frame position.
In particular, the position of output labels in this path corresponds to the frames at which they are most likely emitted.
The probabilities $p^\text{\tiny AV2V}_{\text{\tiny CE}}$ and $p^\text{\tiny AV2A}_{\text{\tiny CE}}$ are derived from the ASR and VSR encoder state sequences by re-using their CTC projection layers from the pre-training stage (i.e., the stage where we train the audio and visual encoders) but here a CE loss is used instead of CTC, because the target alignment $\z^{AV}$ is known. $\beta_{\text{\tiny AV2A}}$ and $\beta_{\text{\tiny AV2V}}$ do control the relative weights for each CE loss term. These auxiliary CE losses are illustrated in Fig.~\ref{fig:architecture}. We should note that the alignments can also be extracted from either the ASR or VSR encoder outputs but instead, we demonstrate that the alignments from joint audio-visual output is optimal (see Section \ref{ssec: response offset}).

\section{Experimental Setup}
\label{sec:experimental setup}
\noindent\textbf{Datasets}\quad
\label{ssec:datasets}
In this work, we use LRS3~\cite{Afouras18d}, which is the largest public benchmarks for audio-visual speech recognition. The dataset is collected from more than 5\,000 speakers in TED and TEDx talks and contains 151\,189 utterances with a total of 438.9 hours. LRS3 is divided into 408 hours of pre-training data, 30 hours of training-validation data, and 0.9 hours of test data.

\noindent\textbf{Pre-processing}\quad
\label{ssec: preprocessing}
For the visual stream, we follow the pre-processing pipeline from~\cite{ma2022visual} to extract the mouth region of interests (ROIs) with a size of 96 $\times$ 96.
Each frame is normalized by subtracting the mean and dividing by the standard deviation of the training set. Audio stream has been normalized by $z$-normalization.

\noindent\textbf{Training details}\quad
\label{ssec: training details}
Similar to \cite{DBLP:journals/corr/abs-2102-06657}, the audio and visual encoder neural networks of the proposed AV-ASR system are using 12 conformer blocks each with an attention dimension of 256, a head size of 8 for the audio stream and 4 for the visual stream, feed-forward dimensions of 2\,048 and a convolution kernel sizes of 31. The number of parameters in the audio front-end, visual front-end and conformer back-end are 3.9\,M, 11.2\,M and 31.8\,M, respectively.
Visual streams are augmented with horizontal flipping, random cropping~\cite{DBLP:journals/corr/abs-2102-06657}, and adaptive time masking~\cite{ma2022visual} while audio streams are augmented with the adaptive time masking. For adaptive time masking, we use one mask per second and for each mask, we randomly mask a duration up to 0.4 second. The audio-only and visual-only models are trained for 75 epochs with AdamW optimizer~\cite{loshchilov2019decoupled} with a dynamic batching strategy, where the maximum number of frames in a batch is set to 2\,400. The learning rate increases linearly with the first 25\,000 steps, yielding a peak learning rate of 0.0006 and thereafter decreases with a cosine annealing strategy.
We use SentencePiece~\cite{DBLP:conf/acl/Kudo18} sub-word units with a vocabulary size of 5\,000.

\noindent\textbf{Algorithmic latency}\quad
The individual encoders of the proposed streaming AV-ASR system are using a total algorithmic delay of 515~ms (audio encoder) and 580~ms (visual encoder), whereby the larger delay is decisive. This delay is composed of the ResNet front-end delay, which amounts to 35 ms (audio) and 100 ms (visual), and the delay caused by the chunk-wise self-attention of the encoder, where we use a total of 12 look-ahead frames which amounts to 480 ms (40~ms frame rate). The triggered attention decoder also uses 12 look-ahead frames which brings to total algorithmic delay to 580~ms (encoder) + 480~ms (decoder) = 1\,060~ms.

\section{Results}
\label{sec:results}
\begin{table}[!t]
\centering
\small
\renewcommand\arraystretch{.85}
\caption{Comparison with state-of-the-art audio-only, visual-only and audio-visual models trained only on the LRS3 dataset. `CM-seq2seq' denotes the conformer-based architecture used in~\cite{DBLP:journals/corr/abs-2102-06657}.}
\vspace{-2mm}
\begin{tabularx}{\linewidth}{l o o o}
\toprule
\textbf{Method} &\textbf{Type} &\textbf{Auxiliary Losses} & \textbf{WER} (\%) \\ \midrule\midrule
\multicolumn{4}{c}{\textit{Non-Streaming Models}} \\ \midrule
CM-seq2seq~\cite{DBLP:journals/corr/abs-2102-06657} &\multirow{4}{*}[-0.1em]{V} &- &46.9 \\
AVHuBERT~\cite{shi2022learning} & &- &41.6 \\
CM-seq2seq~\cite{ma2022visual} & &- &37.9 \\
Ours & &- &\textbf{38.6} \\
\midrule
CM-seq2seq~\cite{DBLP:journals/corr/abs-2102-06657} &\multirow{2}{*}[-0.1em]{A} &- &2.3 \\
Ours & &- &\textbf{2.2} \\
\midrule
CM-seq2seq~\cite{DBLP:journals/corr/abs-2102-06657} &\multirow{3}{*}[-0.1em]{A+V} &- &2.3 \\
\cmidrule(lr){1-1}\cmidrule(lr){3-4}
\multirow{2}{*}[-0.1em]{Ours} & &- &\textbf{2.2} \\
\cmidrule(lr){3-4}
 & &\cmark &\textbf{2.0} \\ \midrule\midrule
\multicolumn{4}{c}{\textit{Streaming Models}} \\ \midrule
\multirow{4}{*}[-0.4em]{Ours} &V &- &\textbf{50.6} \\
\cmidrule(lr){2-4}
 &A &- & \textbf{2.5} \\
\cmidrule(lr){2-4}
&\multirow{2}{*}[-0.1em]{A+V} &- &\textbf{2.6} \\
\cmidrule(lr){3-4}
 & &\cmark &\textbf{2.6} \\ 
\bottomrule
\end{tabularx}
\label{tab:sota lrs3}
\vspace{-2mm}
\end{table}
\noindent\textbf{Non-streaming and streaming models}\quad
Results for our streaming and non-streaming ASR, VSR and AV-ASR models are shown in Table~\ref{tab:sota lrs3}.  Note that our models are trained using only the LRS3 dataset, with a total of 438 hours.
Our non-streaming audio-only and audio-visual models lead to state-of-the-art performance with a WER of 2.2\,\% and 2.0\,\% WER, respectively. Our non-streaming visual-only model achieves slightly lower performance than the existing state-of-the-art model, which uses additional auxiliary losses during training~\cite{ma2022visual}.
Our streaming audio-only model leads to a small increase in the WER of 0.3\,\% compared to the non-streaming model (2.3\,\% WER). On the other hand, the streaming visual-only model results in a large increase in the WER of 12\,\%.
Presumably, the visual-only model is relying on more future context, which is available in the non-streaming case, to resolve visual ambiguities. Finally, the streaming audio-visual model achieves a WER of 2.6\,\% which is 0.6\,\% higher than the non-streaming model. This demonstrates the effectiveness of the proposed model. We should also note that the streaming audio-visual model is slightly worse than the streaming audio-only model mainly due to degraded performance of the streaming visual-only model.

\begin{table}[tb]
\centering
\small
\renewcommand\arraystretch{.87}
\caption{WER [\%] of our models on the LRS3 dataset.}
\vspace{-2mm}
\begin{tabularx}{.99\linewidth}{o o | c c c c c  }
\toprule
\multirow{2}{*}{\textbf{Type}}
&\multirow{2}{*}{\textbf{\shortstack{Auxiliary \\Losses}}}
&\multicolumn{5}{c}{\textbf{SNR Levels [dB]}}
    \\
& & 12.5 & 7.5 & 2.5 & -2.5 & -7.5
    \\
\midrule\midrule
\multicolumn{7}{c}{\textit{Non-Streaming Models}}
    \\
\midrule
A & - & 
3.1 & 5.1 & 12.1 & 38.6 & 89.9
    \\
\midrule
\multirow{2}{*}{A+V} & -  & 
 2.9 & 3.9 & 7.5 & 18.8 &52.9
    \\
\cmidrule(lr){2-7}
& \cmark  & 
 2.7 & 3.7 & 7.8 & 18.4 &43.9
    \\
\midrule\midrule
\multicolumn{7}{c}{\textit{Streaming Models}}
    \\
\midrule

A & - & 
4.2 & 8.4 & 23.6 & 64.9 & 97.9
    \\
\midrule
\multirow{2}{*}{A+V} & - &
4.0 & 7.5 & 16.0 & 38.0 & 74.0
    \\
\cmidrule(lr){2-7}
 & \cmark &
3.8 & 6.8 & 14.9 & 32.4 & 57.5
    \\
\bottomrule
\end{tabularx}
\label{tab: noise experiments}
\vspace{-7mm}
\end{table}

\noindent\textbf{Noise experiments}\quad
\label{ssec: noise experiments}
In order to test the robustness of the AV-ASR models, we add pink noise from the Speech Commands dataset~\cite{DBLP:journals/corr/abs-1804-03209} to the test set at different signal-to-noise ratio (SNR) levels.
For training, babble noise from NOISEX~\cite{DBLP:journals/speech/VargaS93} is added to the clean speech with a random SNR level selected from the set [-5\,dB, 0\,dB, 5\,dB, 10\,dB, 15\,dB, 20\,dB, $\infty$ dB] with a uniform distribution. The results shown in Table~\ref{tab: noise experiments} demonstrate that overall the performance gap between audio-only and audio-visual models is increasing as the level of noise increases.
For example, at the SNR level of -7.5\,dB the streaming audio-visual model results in an absolute improvement of 23.9\,\% WER compared with its audio-only counterpart.
When including alignment regularization by using the auxiliary CE losses, a further reduction of 16.5\,\% WER is achieved.
A similar improvement can also be observed for the non-streaming models. A reduction of 9.0\,\% in WER is achieved at the SNR level of -7.5\,dB for audio-visual non-streaming models when alignment regularization is applied.

\noindent\textbf{Using different modalities for alignment}\quad
\begin{table}[!tb]
\centering
\small
\caption{WER [\%] of our AV-ASR models as a function of noise levels on the LRS3 dataset. We use the alignment information from the \textbf{target modality (T. M.)} and force the response of \textbf{predictive modality (P. M.)} to be closer to it. ``A Frozen'' and ``V Frozen'' denote that the weights of ASR encoder and VSR encoder loaded from the pre-training stage and are kept frozen during fine-tuning the AV-ASR system. ``A Fine-Tuned'' and ``V Fine-Tuned'' denote that the weights of the ASR encoder and VSR encoder are loaded from the pre-training stage and are fine-tuned when training the AV-ASR system.}
\vspace{-2mm}
\renewcommand\arraystretch{.85}
\begin{tabularx}{.99\linewidth}{o o | c c c c c  }
\toprule
\multirow{2}{*}{\textbf{\shortstack{T. M.$\rightarrow$\\P. M.}}}
&\multirow{2}{*}{\textbf{\shortstack{Auxiliary \\Losses}}}
&\multicolumn{5}{c}{\textbf{SNR Levels [dB]}}
    \\
& & 12.5 & 7.5 & 2.5 & -2.5 & -7.5
    \\
\midrule\midrule
\multicolumn{7}{c}{\textit{A Frozen, V Frozen}}
\\
\midrule
- & - & 
4.4&	8.5&	18.2&	38.2&	59.8
    \\
\midrule\midrule
\multicolumn{7}{c}{\textit{A Frozen, V Fine-Tuned}}
    \\
\midrule
- & - &
4.5	 & 8.3	& 19.0	& 43.5	& 80.5
    \\
\midrule
\multirow{1}{*}{A$\rightarrow$V} & \cmark &
4.6	& 8.3	& 17.8	& 35.8	& 56.7
    \\
\midrule\midrule
\multicolumn{7}{c}{\textit{A Fine-Tuned, V Frozen}}
    \\
\midrule
- & - &
4.8	& 8.1	& 17.3	& 39.1	& 69.1
    \\
\midrule
\multirow{1}{*}{V$\rightarrow$A} & \cmark &
4.4	& 7.7	& 16.6	& 34.8	& 56.1
    \\
\midrule\midrule
\multicolumn{7}{c}{\textit{A Fine-Tuned, V Fine-Tuned}}
    \\
\midrule
- & - &
4.0	& 7.5	& 16.0	& 38.0	& 74.0
    \\
\midrule
\multirow{1}{*}[0.0em]{\shortstack{AV$\rightarrow$\{A, V\}}} & \cmark &
3.8	& 6.8	& 14.9	& 32.4	& 57.5
    \\
\bottomrule
\end{tabularx}
\label{tab: aligned loss analysis}
\vspace{-5mm}
\end{table}

In this section, the impact of using different target alignments from different modalities is studied for alignment regularization, i.e., target alignments generated by the audio or visual encoder are compared against using target alignments from the joint audio-visual responses (default case). Results presented in Table~\ref{tab: aligned loss analysis} show that independent of the modality for the target alignment, the overall performance is improved when alignment regularization is applied. For example, when using the alignments from the visual encoder (targeted modality) to supervise the audio counterpart (predictive modality), i.e., V$\rightarrow$A in Table~\ref{tab: aligned loss analysis}, the performance of the AV-ASR model (the fifth row in Table~\ref{tab: aligned loss analysis}) outperforms the model when using the alignments from the audio encoder to supervise the visual counterpart, i.e., A$\rightarrow$V (the third row in Table~\ref{tab: aligned loss analysis}).
It is likely that for A$\rightarrow$V, i.e., forcing the visual stream to temporally align with the audio stream, the output latency is reduced, because the audio encoder tends to respond earlier than the visual encoder (see Table~\ref{tab: response offset analysis}). 
However, the best WER are obtained when using target alignments generated from the joint AV output.

\noindent\textbf{Response offset analysis}\quad
\label{ssec: response offset}
\begin{table}[tb]
\centering
\small
\caption{The mean response offset over the test set of LRS3 using a streaming AVSR model.
``V$\rightarrow$AV'' with a positive $\textbf{N}$ denotes the visual encoder outputs are relatively slower to the the joint audio-visual representations with $\textbf{N}$ frames, and vice versa.}
\vspace{-2mm}
\renewcommand\arraystretch{.85}
\begin{tabularx}{.99\linewidth}{o o | c c c c c  }
\toprule
\multirow{2}{*}{\textbf{\shortstack{Auxiliary \\Losses}}}&
\multirow{2}{*}{\textbf{\shortstack{Relative \\Offset}}} &\multicolumn{5}{c}{\textbf{SNR Levels [dB]}}
    \\
& & 12.5 & 7.5 & 2.5 & -2.5 & -7.5
    \\
\midrule\midrule
\multicolumn{7}{c}{\textit{A Frozen, V Frozen}}
    \\
\midrule
\multirow{2}{*}{-} & V$\rightarrow$AV &
1.11&	1.02&	0.86&	0.57&	0.20
    \\
 & A$\rightarrow$AV &
0.01&	0.02&	0.02&	0.22&	5.68
    \\
\midrule\midrule
\multicolumn{7}{c}{\textit{A Fine-Tuned, V Fine-Tuned}}
    \\
\midrule
\multirow{2}{*}{-} & V$\rightarrow$AV &
0.22&	0.18&	0.11&	0.07&	0.31
    \\
 & A$\rightarrow$AV &
0.00&	0.01&	0.04&	0.23&	-3.34
    \\
\midrule\midrule
\multicolumn{7}{c}{\textit{A Fine-Tuned, V Fine-Tuned}}
    \\
\midrule
\multirow{2}{*}{\cmark} & V$\rightarrow$AV &
0.14&	0.10&	0.04&	0.02&	0.03
    \\
 & A$\rightarrow$AV &
0.00&	0.01&	0.04&	0.20&	1.60
    \\
\bottomrule
\end{tabularx}
\label{tab: response offset analysis}
\vspace{-4mm}
\end{table}

In order to better understand how the proposed auxiliary tasks contribute to learning better alignments between audio and visual streams, we analyze the response offset between the joint audio-visual representations and either the audio or visual encoder outputs for the proposed streaming AV-ASR models. In particular, we calculate the average frame distance between the forced alignment sequences $z^{AV}$ and either $z^A$ or $z^V$ for the ASR labels over the entire test set of LRS3. Results are shown in Table~\ref{tab: response offset analysis}. It can be observed that without alignment regularization, the joint AV response tends to be better aligned with the audio modality at high SNR levels and only for low SNR levels with the visual modality. This indicates that the modality which is better aligned with the joint audio-visual representations is more likely responsible for the final predictions. In other words, the modality with larger delay relative to the joint audio-visual representations is more likely to be ignored. This property is consistent with the results reported in Table~\ref{tab: noise experiments}, i.e., the performance of streaming audio-visual models outperforms the streaming audio-only counterpart especially at high SNR levels. Table~\ref{tab: response offset analysis} also demonstrates that fine-tuning the audio and visual encoders leads to better alignments between both modalities and better ASR results as shown in Table~\ref{tab: aligned loss analysis}. However, when using alignment regularization in training, we observe that the maximum response offset between the VSR encoder and the joint audio-visual representations is less than 0.14 frames (5.6 ms) across all SNR levels. This indicates that a better synchronization between ASR and VSR encoder state sequences leads to a more robust performance~(the last row in Table~\ref{tab: aligned loss analysis}), especially when the audio stream is heavily corrupted by noise.

\section{Conclusion}
\label{sec:conclusion}
In this work, we presented an end-to-end streaming AV-ASR system based on the conformer architecture. We showed that visual-only models are performing substantially worse as compared to the audio-only counterpart when comparing their streaming and non-streaming results, probably due to the assumption that visual streams relies more on the future context. Furthermore, we showed that by adding alignment regularization, the performance of streaming and non-streaming AV-ASR models is much improved with an increasing impact for higher noise levels. The response time between the audio and visual modalities is analyzed and we demonstrated that:~1) The modality with smaller latency relative to the joint audio-visual representations is more responsible to the final predictions.~2) The proposed alignment regularization technique results in a better response synchronization between the audio and visual encoder streams leading to improved joint audio-visual predictions and an increased robustness.

\clearpage
\section{References}
\begingroup
\printbibliography[heading=none]
\endgroup

\end{document}